\documentclass[twocolumn,times]{aastex631}

\usepackage[T1]{fontenc}

\newcommand\ergs{erg~s$^{-1}$}

\usepackage{soul}
\usepackage{amsmath}

\shorttitle{A Neutron Capture Explanation for the 10 MeV Line in GRB 221009A}
\shortauthors{Zhu et al.}

\graphicspath{{./}{fig/}}

\begin{document}

\title{A Neutron Capture Explanation for the 10 MeV Emission Line Seen in GRB 221009A}

\author[0000-0002-9337-9974]{Jiahuan Zhu}
\affiliation{Department of Astronomy, Tsinghua University, Beijing 100084, China}

\author[0000-0001-7584-6236]{Hua Feng}
\affiliation{Key Laboratory of Particle Astrophysics, Institute of High Energy Physics, Beijing 100049, China}
\email{hfeng@ihep.ac.cn}


\author[0000-0001-8678-6291]{Tong Liu}
\affiliation{Department of Astronomy, Xiamen University, Xiamen, Fujian 361005, China}

\begin{abstract}

The brightest ever gamma-ray burst (GRB) 221009A displays a significant emission line component around 10 MeV. 
As the GRB central engine is neutron-rich, we propose that the emission line could be originally due to the 2.223 MeV gamma-rays following neutron capture with protons. 
The measured line profile can be adequately fitted with a neutron capture model that involves thermal broadening and a bulk Doppler shift. 
The spectral modeling reveals a Doppler factor varying from 5.1 to 2.1 for the neutron-rich component, along with a temperature increase from 300~keV to about 900~keV, during the time interval of 280--360 s since the trigger, with about $10^{-2}$~$M_\sun$ deuteriums produced in the process. 
We argue that the neutron capture can take place in the outer shell of a structure jet. Disk winds could be another possible site.

\end{abstract}


\section{Introduction} 
\label{sec:intro}

The long-duration gamma-ray burst (GRB) 221009A is the brightest GRB observed so far \citep{2023arXiv230301203A, 2023ApJ...946L..31B, 2023ApJ...949L...7F}. It displayed the highest total isotropic energy, $E_{\gamma,{\rm iso}} = 1.5 \times 10^{55}$~erg, and the second highest isotropic peak luminosity, $L_{\gamma,{\rm iso}} = 1.7 \times 10^{54}$~\ergs\ \citep{2023arXiv230301203A}. The burst is followed by a faint supernova SN 2022xiw \citep{2023ApJ...949L..39S} in a galaxy at a redshift  $z = 0.151$ or a luminosity distance of 745~Mpc \citep{2023arXiv230207891M}. One of the most intriguing discoveries associated with this burst is a significant detection of very high energy (VHE) gamma-rays with LHAASO shortly after the trigger, with the highest photon energy above 10~TeV \citep{2023SciA....9J2778C}.

Analysis of the data obtained with GECAM, an instrument that did not suffer from saturation during the peak of the burst, suggests that the main emission and flares are mainly due to synchrotron radiation from relativistic electrons \citep{2023ApJ...947L..11Y}. Combining the data of GECAM, Fermi, and LHAASO, \citet{2023arXiv231011821W} argued that both leptonic and hadronic processes are needed to reproduce the observed spectra and lightcurve. 

\citet{ZhengC2024} found an achromatic jet break and energy-dependent flux decays in the early afterglow, in support of the scenario of a structured jet with a bright narrow core. \citet{2023MNRAS.522L..56S} argued that the afterglow emission from radio to VHE gamma-rays could be explained with a two-component jet model, consisting of a narrow jet with a higher Lorentz factor and a wide jet with a lower one. \citet{Zhang2024} also proposed a two-component jet model, with a narrow Poynting-flux-dominated jet surrounded by a matter-dominated structured jet wing, which can account for a variety of observational features of the burst; \citet{2023arXiv231113671Z} further showed that such a two-component jet model with forward and reverse shocks could explain both the energy flux and spectral evolution seen in the VHE band. By modeling the TeV and multiwavelength afterglows,  \citet{Zheng2024} found that the emission can be interpreted by a narrow uniform core and a wide structured wing, with a small Lorentz factor for the outer one.

After ruling out possible instrumental effects, \citet{Ravasio2024} claimed the detection of a varying emission line at an energy of around 10~MeV in this burst with the BGO detector onboard the Fermi Gamma-ray Burst Monitor (GBM). They fitted the energy spectra with a model including a smoothly broken power-law (SBPL) component for the continuum and a Gaussian component for the emission line, and revealed a significant $\sim$10~MeV line during an unsaturated time interval from $t_0 + 280$~s to $t_0 + 360$~s, where $t_0$ is the GBM trigger time on 2022 October 9 at 13:16:59.99 UTC \citep{2023ApJ...952L..42L}. The centroid energy of the emission line was found to evolve with time, varying from about 12~MeV in the epoch of 280--300~s to about 6~MeV in 340--360~s since the trigger, along with a flux decay by a factor of $\sim$5. 
No significant variation can be found in the line width, which is around 1--2~MeV FWHM, indicative of a relatively narrow emission line. 
\citet{Burns2024} excluded instrumental effects and determined a significance over 5$\sigma$ for the line emission.
With unsaturated GECAM data, \citet{ZhangYQ2024} restored part of the saturated GBM data, which allows them to characterize the emission line back to $t_0 + 246$~s; they found that the central energy and line flux decays with a power-law relation $t^{-1}$ and $t^{-2}$, respectively.


The feature was interpreted as a blue-shifted annihilation line from relatively cold ($kT < m_{\rm e} c^2$) electron-positron pairs with a moderate Lorentz factor $\Gamma \sim 20$. \citet{Ravasio2024} argued that the line emitting region was a result of collision between a very fast jet shell and a slow one, producing an optically thick shell for pair production. 
Considering the central jet within 1$\arcdeg$, \citet{ZhangZ2024} found the central energy evolution could be naturally explained by the high-latitude curvature effect but the line flux decay deviated from the expected relation of $t^{-3}$. 
\citet{Peer2024} argued that electron-position pair annihilation at high-latitude can naturally explain the line evolution.
The narrow line width of 10\% requires fast cooling, which may imply a magnetic field-dominated jet \citep{ZhangZ2024}. 
\citet{Yi2024} established a self-consistent picture of electron-positron production, cooling, and annihilation, constraining the regions where the pairs were produced and annihilated.
Alternatively, \citet{Wei2024} proposed that the emission line was due to fluorescent emission from hydrogen-like heavy ions. 

The GRB central engine is expected to be hot, dense, and neutron-rich \citep{Derishev1999, Beloborodov2003}. 
The accretion disk has a high neutron fraction \citep{Beloborodov2008}, and the neutron-rich material will enter the relativistic jet \citep{Levinson2003, Metzger2008} due to coupling with protons via elastic collisions \citep{Beloborodov2010}. 
Therefore, here in this paper, we propose an alternative interpretation of the 10~MeV emission line seen in GRB 221009A as due to neutron capture (\S~\ref{sec:model}). 
We fit the energy spectra with the neutron capture model (\S~\ref{sec: fitting}) and discuss the possible physical origins under this scenario (\S~\ref{sec:discussion}).

\section{Model}
\label{sec:model}

When a near-rest neutron is captured by a near-rest proton, a deuterium forms and releases a gamma-ray photon with the binding energy (2.223~MeV) of deuterium,
\begin{equation}
\rm p + n \rightarrow D + \gamma~(2.223~MeV) \, .
\label{eq:neutron_capture}
\end{equation}

The cross section of this process and the energy of the gamma-ray photon depend on the kinematics of the particles, which also leads to spectral broadening and shift. 
Assuming a non-relativistic plasma and a Maxwellian kinetic energy distribution, the observed gamma-ray spectrum given a bulk of neutron capture can be described as \citep{1984MNRAS.210..257A}
\begin{equation}
\Phi(E_\gamma)=A_0\left(\frac{E_\gamma-W_1}{kT}\right)^{3/2}\frac{\exp[-(E_\gamma-W_1)/kT]}{E_\gamma} \, ,
\label{eq:rest-frame}
\end{equation}
where $A_0$ is the normalization associated with the deuterium density or number, $W_1$ is the deuterium binding energy (2.223~MeV), $T$ is the temperature of particles, $k$ is the Boltzmann constant, and $E_\gamma$ is the observed photon energy in the rest frame.

We assume that the neutron capture emission arises from a conical component with a half-opening angle $\theta$ surrounding the narrow jet. 
If the line of sight coincides with the direction of the jet, the observed energy and flux are both enhanced due to Doppler boosting, by a factor of $D_{\rm n}$ and $D_{\rm n}^3$, respectively, where $D_{\rm n}$ is the Doppler factor
\begin{equation}
D_{\rm n} = \frac{1}{\Gamma_{\rm n}(1-\beta \cos\theta)} \, ,
\label{eq:doppler-factor}
\end{equation}
determined by the Lorentz factor $\Gamma_{\rm n}$ ($=1/\sqrt{1 - \beta_{\rm n}^2}$) or the velocity $\beta_{\rm n}$, and the viewing angle $\theta$. The temperature $T$, Doppler factor $D_{\rm n}$, and normalization $A_0$ can be obtained with spectral fits.

\section{Spectral fits} 
\label{sec: fitting}

Following \citet{Ravasio2024}, we extracted the BGO energy spectra in the 0.3--35~MeV band and in the same four epochs, i.e., 280--300~s, 300--320~s, 320--340~s, and 340--360~s since the trigger. The response matrix files (rsp2) were retrieved from HEASARC\footnote{\url{https://heasarc.gsfc.nasa.gov/FTP/fermi/data/gbm/triggers/2022/bn221009553/current/}} and interpolated to the times of interest. We adopted the CSPEC data with the 1024 ms time resolution from BGO 0 and 1 detectors. The background was estimated from the data before ($t_0 - 50$~s to $t_0 - 5$~s) and after ($t_0 + 700$~s to $t_0 + 750$~s) the burst, with a first-order polynomial fitting and interpolation in the four epochs using the \textit{interpolate\_bins} command. The time-integrated background spectra were then obtained using the \textit{integrate\_time} command. The \textit{grppha} tool was used to group the spectra linearly by a factor of 3.

\begin{figure*}
\includegraphics[width=0.95\linewidth]{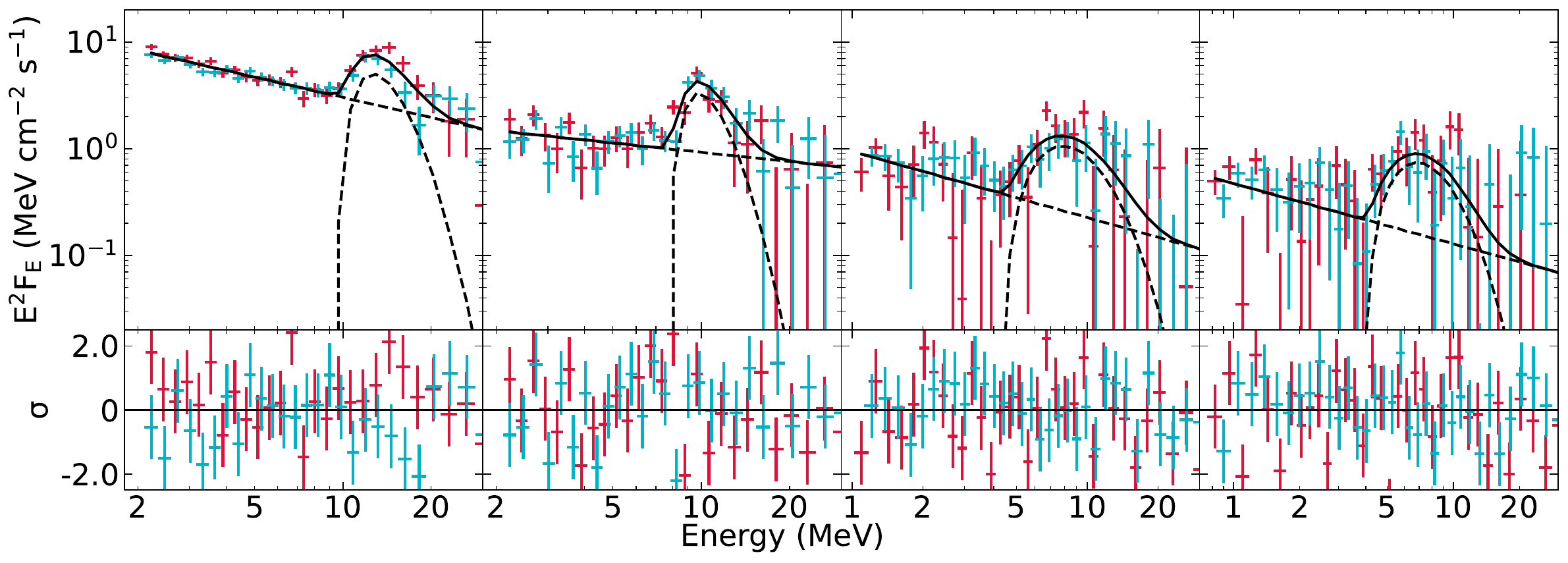}
\caption{Fermi/GBM BGO spectral fitting with a power-law plus neutron capture model in four epochs, respectively from left to right, in the time interval of 280--300~s, 300--320~s, 320--340~s, and 340--360~s since the trigger of GRB 221009A. The red and blue points represent the data from the two BGO detectors, b0 and b1, respectively.}
\label{fig:spectra}
\end{figure*}

We adopted the power-law model for the continuum and used the neutron capture model to account for the line feature, both modified by the cosmological redshift with the {\tt zashift} model in XSPEC. Based on the spectral results in \citet{Ravasio2024}, we chose a low energy cut of 2, 2, 1, and 0.7~MeV, respectively, for the spectra in the four epochs, to discard the data below the spectral break. We note that the fitting results remain consistent within errors for the line feature if we use a lower energy threshold and adopt the SBPL model for the continuum, while the simple power-law model allows for a more robust determination of the line feature, less affected by imperfect modeling of the continuum. The energy spectra and best-fit models are shown in Figure~\ref{fig:spectra}, and the best-fit spectral parameters are listed in Table~\ref{tab:fit}. Here we derive the total line count rate assuming a half opening angle of 10$\arcdeg$. As one can see, the neutron capture model provides adequate fits to the data.

\tabletypesize\scriptsize
\begin{deluxetable*}{cccccccc}
\caption{Best-fit parameters for the neutron capture model in the four epochs.}
\label{tab:fit}
\tablehead{
\colhead{Epoch} & \colhead{PhoIndex} & \colhead{$D_{\rm n}$} & \colhead{$kT$} & \colhead{$R_{\rm obs}$} & \colhead{$F_{\rm obs}$} & \colhead{$R_{\rm int}$} & \colhead{$\chi^2$/d.o.f.} \\
\colhead{(s)} & \colhead{} & \colhead{} & \colhead{(keV)} & \colhead{(cts~$\rm cm^{-2}$~s$^{-1}$)} & \colhead{($10^{-6}$~erg~$\rm cm^{-2}$~s$^{-1}$)} & \colhead{($10^{52}$~cts~s$^{-1}$)} & \colhead{} \\
\colhead{(1)} & \colhead{(2)} & \colhead{(3)} & \colhead{(4)} & \colhead{(5)} & \colhead{(6)} & \colhead{(7)} & \colhead{(8)}
}
\startdata
 280--300 & $2.63^{+0.08}_{-0.08}$ & $5.07^{+0.42}_{-0.45}$ & $364.2^{+154.9}_{-116.9}$ & $0.12^{+0.02}_{-0.02}$ & $1.93^{+0.37}_{-0.36}$ & $1.41^{+0.27}_{-0.27}$ & 55.0/51 \\ 
 300--320 & $2.28^{+0.69}_{-0.21}$ & $4.11^{+0.50}_{-1.50}$ & $294.2^{+840.7}_{-170.1}$ & $0.09^{+0.08}_{-0.02}$ & $1.14^{+0.96}_{-0.30}$ & $1.27^{+0.11}_{-0.33}$ & 69.9/51 \\ 
 320--340 & $2.59^{+0.45}_{-0.28}$ & $2.31^{+0.72}_{-0.61}$ & $872.4^{+765.1}_{-522.2}$ & $0.07^{+0.03}_{-0.03}$ & $0.53^{+0.22}_{-0.20}$ & $1.88^{+0.78}_{-0.70}$ & 62.2/62 \\ 
 340--360 & $2.53^{+0.53}_{-0.33}$ & $2.06^{+0.74}_{-0.72}$ & $817.6^{+985.6}_{-614.1}$ & $0.06^{+0.03}_{-0.03}$ & $0.37^{+0.17}_{-0.17}$ & $1.63^{+0.74}_{-0.77}$ & 77.5/66 \\ 
\enddata

\tablecomments{
(1) Observed time interval since the trigger. 
(2) Rest-frame power-law photon index. 
(3) Doppler factor. 
(4) Rest-frame gas temperature. 
(5) Observed emission line count rate. 
(6) Observed emission line flux.
(7) Rest-frame emission line count rate assuming a half opening angle of 10$\arcdeg$.
(8) best-fit $\chi^2$ and degree of freedom. 
Errors are quoted at the 90\% confidence level.}
\end{deluxetable*}

\section{Discussion} 
\label{sec:discussion}

We demonstrate that the 10~MeV line feature seen in the late-time (280--360~s) energy spectra of GRB 221009A can be adequately described with a neutron capture model. 
Here we briefly describe the overall physical picture and discuss the related physical properties. 

In the GRB jet, neutrons first travel together with protons at the same $\Gamma$ via elastic collisions until a radius $r_{\rm n}$, where the timescale for n-p collisions becomes longer than the jet expansion timescale. 
Beyond $r_{\rm n}$, protons and neutrons decouple; protons keep accelerating while neutrons start to coast \citep{Bahcall2000}, forming compound flows in which protons move faster than neutrons \citep[$\Gamma > \Gamma_{\rm n}$;][]{Beloborodov2010}.
The decoupling may happen either with a large Lorentz factor \citep{Bahcall2000}, e.g., $\Gamma \gtrsim 400$, or due to large variations of the Lorentz factor, i.e., strong jet variability \citep{Beloborodov2010}.
After the formation of compound flows, the jet becomes transparent to radiation at the photosphere $r_\star$. Also, due to the short lifetime of neutrons, they can only travel to the decay radius $r_\beta$. 
In general, one has $r_{\rm n} < r_\star < r_\beta$ in the GRB jet \citep{Beloborodov2010}. In this picture, the neutron capture line is only detectable if the capture happens between the $ r_\star$ and $r_\beta$.

\subsection{Dynamics of the neutron-rich component}
{\label{sec:structured-jet}

According to Eq.~(\ref{eq:doppler-factor}), the Lorentz factor $\Gamma_{\rm n}$ and opening angle $\theta$ are degenerate. 
Given the best-fit $D_{\rm n}$, the relations between $\Gamma_{\rm n}$ and $\theta$ in the four epochs are shown in Figure~\ref{fig:theta-gamma}. 
The opening angle of the narrow jet is estimated as $\sim0.8^\circ$ \citep{ 2023Sci...380.1390L, 2023arXiv230301203A}, which sets the lower limit of $\theta$. 

We adopt the structured jet model \citep{Zhang2024}, which, in principle, does not consider the neutron component and its decoupling with protons at late time. Therefore, the model can only be regarded as an approximation in this case.
Based on the model, the $\Gamma_{\rm n}$ vs.\ $\theta$ relation is determined by the parameter $k_{\Gamma_{\rm n}}$. 
One has $\Gamma_{\rm n} \propto \theta^{-1}$ for $k_{\Gamma_{\rm n}} \leq 1$ or $\Gamma_{\rm n} \propto \theta^{-k_{\Gamma_{\rm n}}}$ for $k_{\Gamma_{\rm n} >1}$. 
Then, both $k_{\Gamma_{\rm n}}$ and the parameter $k_\epsilon$ (about the angular distribution of energy) determine the exact $\Gamma_{\rm n}$ and $\theta$ at a given time. 
We searched in the allowed parameter space, and displayed an example with $k_{\Gamma_{\rm n}}=1.1$ and $k_\epsilon=2$ (dotted line in Figure~\ref{fig:theta-gamma}), in which case the predicted $\Gamma_{\rm n}$ and $\theta$ coincide with observations in the time interval from $t_0 + 280$~s to $t_0 + 360$~s (the thick segment along the dotted line), suggesting that the structured jet could be the neutron capture site from a dynamical point of view.

In this case, the neutron capture likely occurs in an outer shell of a structured jet, which decelerates from $\Gamma_{\rm n} \approx 40$ to about 10, and the opening angle enlarges from about 5$\arcdeg$ to 15$\arcdeg$. Along with the deceleration, there is concurrent heating ($kT$ increases from about 300~keV to about 900~keV) along with a line flux drop, likely due to interaction with the ambient medium and consequently conversion of the kinematic energy into internal energy. 

Alternatively, the very strong radiation pressure of the accretion disk might create relativistic winds with large-scale magnetic fields and accelerate particles to velocities $\beta \approx 0.2 - 0.9$, lasting hundreds of seconds for long-duration GRBs \citep{2018ApJ...852...20L}. 
Therefore, disk winds could be another possible site where neutron capture occurs; a detailed model is needed to examine whether or not its evolution on the $\Gamma_{\rm n}-\theta$ diagram is consistent with observations. 

\begin{figure}
\includegraphics[width=\columnwidth]{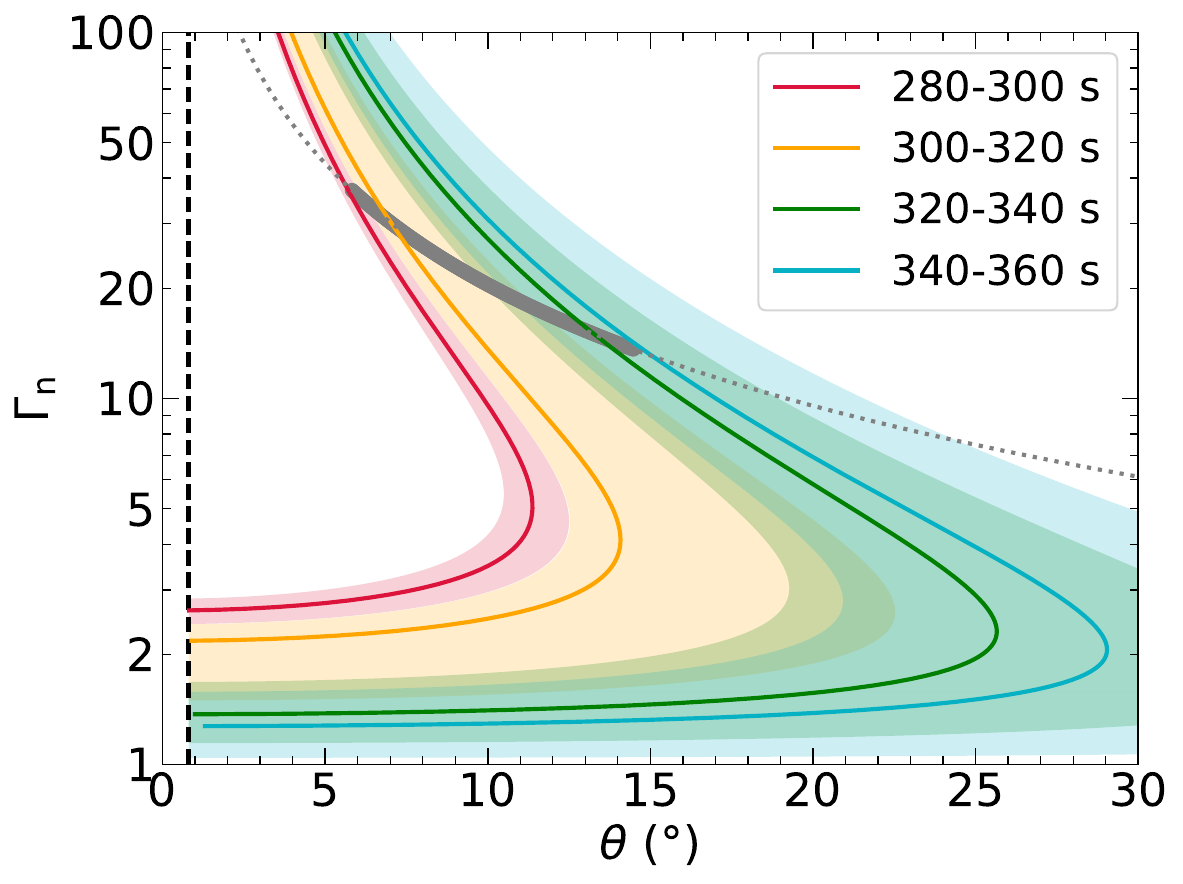}
\caption{Lorentz factor vs.\ half opening angle of the neutron-rich ejecta given the best-fit Doppler factor in the four epochs. The shaded areas represent the 90\% uncertainty range for each curve. The vertical dashed line marks the opening angle of the central jet at 0.8\arcdeg, which is the lower limit of $\theta$. The dotted line represents a possible solution in the structured jet model \citep{Zhang2024}; the gray thick segment highlights the model predicted $\Gamma_{\rm n}$ and $\theta$ during the four epochs.}
\label{fig:theta-gamma}
\end{figure}


\subsection{Reaction rate of the neutron capture process}
\label{sec:reaction-rate}

We performed the nucleosynthesis simulations with \textit{net.Evolve} in SkyNet \citep{Lippuner2017} in the time interval of 245--400~s. We assumed an initial density typical of the density of the GRB disk as $10^{8}$~g~cm$^{-3}$~\citep{Liu2017} and a free expansion ($\propto t^{-3}$) followed by a jet deceleration ($\propto t^{-1.5}$) at the peak of the afterglow (245~s in the observer's frame~\citep{2023Sci...380.1390L}). 
We assumed a constant temperature of 6~GK such that the output temperature roughly match the observed, and also $\Gamma = 20$, a typical Lorentz factor shown in Figure~\ref{fig:theta-gamma}.
The simulation results are shown in Figure~\ref{fig:sim} including the temperature, mass abundance, and reaction rate as a function of time.

\begin{figure}
\includegraphics[width=\columnwidth]{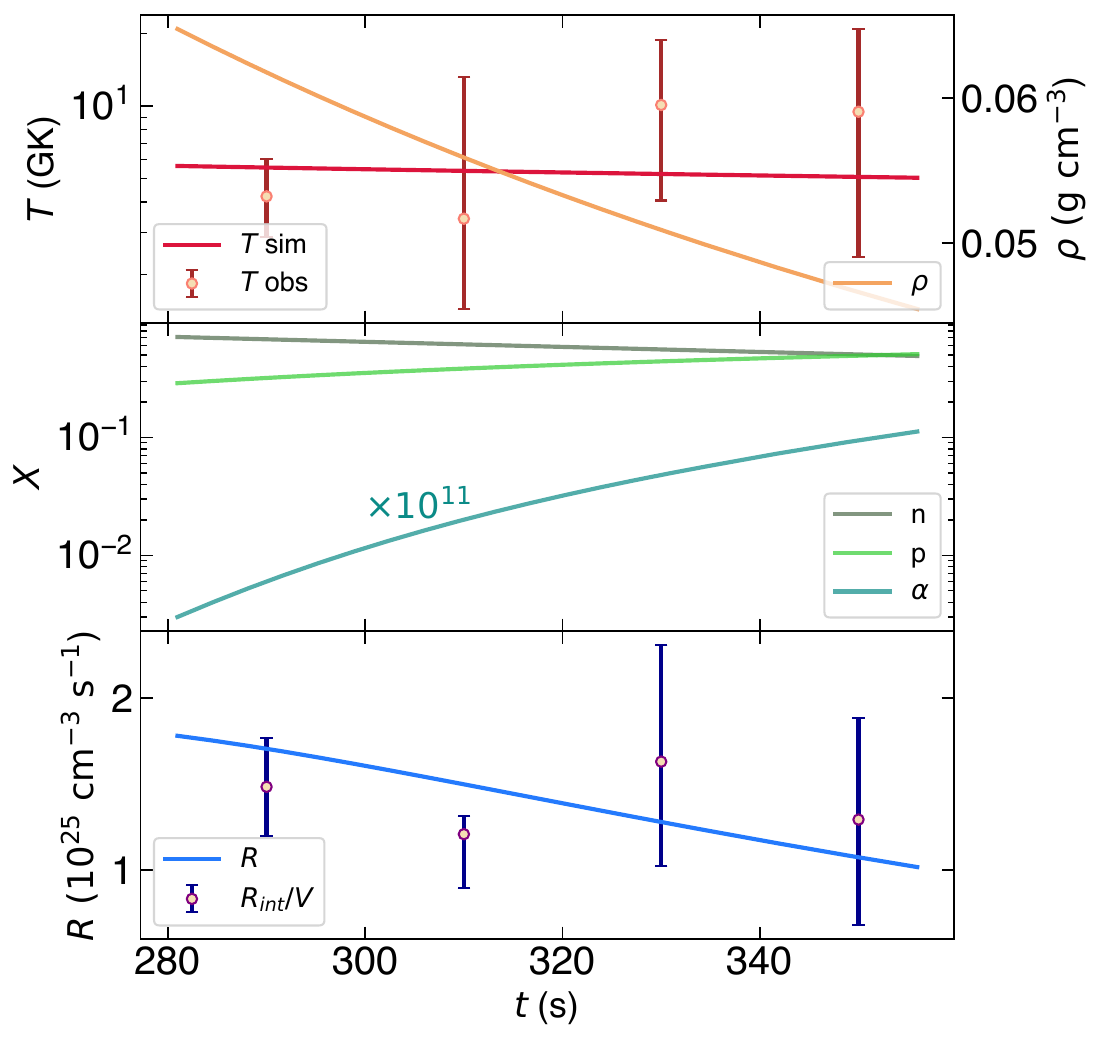}
\caption{Nucleosynthesis simulation results as a function of time and comparison with observations. {\bf Top}: assumed density and output temperature  as a function of time, compared with observed temperatures. {\bf Middle}: evolution of the mass abundance for neutrons, protons, and alpha particles. {\bf Bottom}: evolution of the reaction rate against observations assuming a volume of $1\times10^{27}$~cm$^3$.}
\label{fig:sim}
\end{figure}

The neutron capture reaction rate can be calculated as $\lambda = N_{\rm A} \langle \sigma v \rangle$~[cm$^3$~mol$^{-1}$~s$^{-1}$], where $\sigma$ denotes the cross-section, $v$ the velocity, and $N_{\rm A}$ the Avogadro constant.  The $\langle \sigma v\rangle$ values are retrieved from the JINA REACLIB database  \citep{Cyburt2010}. By combining with the neutron and proton number densities, we computed the volumetric reaction rate as $R = \lambda (\frac{\rho X_{\rm p}}{A_{\rm p}}) (\frac{\rho X_{\rm n}}{A_{\rm n}})N_{\rm A}$~[cm$^{-3}$~s$^{-1}$], where $\rho$ is the mass density, $X_{\rm p}$ and $X_{\rm n}$ represent the mass fractions of protons and neutrons, respectively, and $A_p$ (=1.007 g~mol$^{-1}$) and $A_{\rm n}$ (=1.009 g~mol$^{-1}$) are their corresponding molar masses. The derived reaction rates are displayed in Figure~\ref{fig:sim}. 
To reconcile the simulation with observation, one needs to assume a reaction volume of $(1.0 - 1.3) \times 10^{27}$~cm$^3$ in the rest frame. The physical size of the reaction volume can be calculated as $V^{1/3}\gamma(1+\beta)\sim 4\times10^{10}$~cm. According to~\citet{Gao2015}, the photospheric radius of GRB 221009A is calculated to be $10^{11}-10^{12} $~cm. Under an opening angle of 2$^\circ$, the corresponding size becomes $10^{9}-10^{10}$~cm, while the reaction volume size exceeds this range, demonstrating consistency. The size calculated via $\beta c\Delta t/(1+z)\sim10^{11}$~cm remains comparable with simulations. Thus, the simulation results are consistent with observations.[Discuss if this is consistent with observations. ]


Another way to check the self-consistency of the model is to see if the total mass and mechanical power needed for neutron capture are possible in the scenario of a GRB jet. 
Given the inferred line count rate and epoch duration $\Delta t$, the total mass of deuteriums can be calculated as $R_{\rm int} \Delta t \, m_{\rm D} D_{\rm n} / (1 + z)$, where $m_{\rm D}$ is the deuterium mass and $R_{\rm int}$ is the rest-frame emission line count rate.
Plugging in the measurements quoted in Table~\ref{tab:fit}, we obtained a total deuterium mass of 2.2, 1.6, 1.3, $1.0 \times 10^{-3}$~$M_\sun$, respectively, in the four epochs. 
Then, we can estimate the mechanical power of the neutron-rich component as $P = R_{\rm int} m_{\rm n} c^2 D_{\rm n}^2/ (1+z)(1-Y_{\rm e})$, where $m_{\rm n}$ represents the rest mass of a neutron. 
Assuming $Y_{\rm e} = 0.15$ \citep{Liu2017}, the power is 4.8, 2.9, 1.3, and $0.9 \times~10^{50}$~\ergs\ in the four epochs, roughly following a $t^{-1.4}$ relation, which is also consistent with predictions from the structured jet model \citep{Zhang2024}.


\subsection{Origin of neutrons}
\label{sec:neutron}

The next question is whether or not there are sufficient neutrons. 
The GRB central engine or the accretion disk is neutron-rich due to electron/antineutrino capture on protons, and could also be neutron excessive when the accretion rate is sufficiently high \citep{Beloborodov2003,Metzger2008,Liu2017}. 
The electrons become degenerate when the mass accretion rate is extremely high, in which case the chemical potential $\mu$ is larger than $\phi={kT} / {m_{\rm e}c^2}$. Degeneracy exponentially suppresses the positron density, ${n^+} / {n^-} \propto e^{-\mu / \phi}$, based on the Fermi-Dirac distribution, and thus suppresses the neutron-to-proton process. If the accretion rate is higher than 0.1~$M_\sun$~s$^{-1}$, such an accretion state can last for tens of seconds. Even if the accretion rate is as low as 0.001~$M_\sun$~s$^{-1}$, the inner part of the accretion disk is still neutron-rich and the state can last for hundreds of seconds. Therefore, it is possible that neutrons can exist for an extended period of time. 
This leads to nucleosynthesis and produces a significant amount of deuteriums, which is found to occupy several percent of the total mass and be the most abundant element next to free nucleons and $\alpha$-particles \citep{Beloborodov2003}. 
Therefore, forming $6 \times 10^{-3}$~$M_\sun$ of deuteriums is possible following the collapse of a massive star. 


\subsection{Optical depth and visibility}
\label{sec:visibility}

The line emitting region should be optically thin for the gamma-rays to escape. 
The optical depth of the jet at radius $r$ can be calculated as $\tau = n_{\rm e} \sigma_{\rm T} r / D_{\rm n}$, where $n_{\rm e}$ is the number density of electrons and $\sigma_{\rm T}$ is the Thomson cross-section. 
Assuming that the number densities of electrons, protons, and neutrons are of the same order, the isotropic luminosity of the jet is $L = 8 \pi r^2 \Gamma^2 m_{\rm p} c^3 n_{\rm e}$. We note that neutrons and protons are decoupled at the time of capture, and $\Gamma$ here refers to the Lorentz factor of protons/electrons. 
The optical depth can be calculated as $\tau = L \sigma_{\rm T} / 8 \pi r m_{\rm p} c^3 \Gamma^2 D_{\rm n}$ and the photosphere where $\tau = 1$ is located at $r_{\rm \star} = L \sigma_T / 8 \pi m_{\rm p} c^3 \Gamma^2 D_{\rm n}$. 
In the observer's frame, the time for the jet to reach the photosphere is $t_\star = r_\star / D_{\rm n}^2 c = L \sigma_T /8 \pi m_{\rm p} c^4 D_{\rm n}^3 \Gamma^2$.

When the emission line is detected at time $t$, one has $t > t_\star$ and  $\Gamma^2 > L\sigma_{\rm T}/8 \pi m_{\rm p} c^4 D_{\rm n}^3 t$. Given the measurements, we obtain $\Gamma > 45$ at $t_0 + 280$~s with $L\approx3.83\times10^{51}$~\ergs\ \citep{Ravasio2024}. 
\citet{2023ApJ...956L..38G} estimated a Lorentz factor $\Gamma > 50$ at $t_0 + 280$~s, consistent with our estimate.
Compared with the gray thick segment in Figure~\ref{fig:theta-gamma}, the Lorentz factor of the neutron component is $\Gamma_{\rm n}\approx40$ at $t_0 + 280$~s, satisfying $\Gamma > \Gamma_{\rm n}$. 
\citet{ZhangYQ2024} detected this emission line as early as $t_0 + 246$~s at a higher energy (37~MeV). 
Based on the evolution trend that they inferred, we estimate a Doppler factor $D_{\rm n} \sim 17$ and a jet luminosity $L \approx 10^{52}$~\ergs, and consequently require the Lorentz factor $\Gamma > 13$, which is not constraining.



However, we mention the caveat that the James Webb Space Telescope (JWST) near-infrared observations did not find evidence for r-process products associated with GRB 221009A \citep{Blanchard2024}. Future analysis may translate our scenario to an upper limit of the r-process signature in the JWST data, to further check the model.

\begin{acknowledgments}
We thank the anonymous referee for useful comments, and also Bing Zhang, Xilu Wang and He Gao for helpful discussions. HF acknowledges funding support from the National Natural Science Foundation of China under grants Nos.\ 12025301 \& 12103027, and the Strategic Priority Research Program of the Chinese Academy of Sciences.
\end{acknowledgments}

\vspace{5mm}
\facilities{Fermi (GBM)}

\bibliography{grb}{}
\bibliographystyle{aasjournal}

\end{document}